\documentclass[12pt]{article}

\renewcommand{\footnote}[1]{
\def\thefootnote{\arabic{footnote})}
\footnotemark
\footnotetext{#1}}

\newcommand{\ds}[1]{{\displaystyle{#1}}}
\newcommand{\ts}[1]{{\textstyle{#1}}}

\def\text{\;\;\;\;}
\def\Jnf{J_{|n-\Phi|}}
\def\f{\Phi}
\def\vx{{\bf{x}}} 

\def\vp{{\bf{p}}} 
\def\wwd{{\frac{d}{2}}}
\def\wwdd{{\frac{d}{2}+1}}

\newcommand{\bfnabla}{\mbox{\boldmath $\nabla$}}
\newcommand{\bfdd}{\mbox{\boldmath $\partial$}}
\newfont{\blackboard}{msbm9 scaled\magstep2}
\newfont{\fff}{msbm9 scaled\magstep2}

\newcommand{\Z}{\mbox{\blackboard\symbol{"5A}}}
\def\ugl{\varphi} 
\def\ups{\upsilon} 
\def\dd{\partial} 
\def\ch{{\rm ch}}
\def\sh{{\rm sh}}
\def\Arch{{\rm Arch}}
\renewcommand{\Re}{{\rm Re}}

\def\G{\Gamma}
\def\l{\lambda}
\newcommand{\Sum}[1]{\displaystyle{\sum_{#1} \kern-1.20em \int}} 

\def\Res{\mathop{\rm Res}}

\def\integ{\int_0^{\infty}}

\def\begineq{\begin{eqnarray}}
\def\endeq{\end{eqnarray}}

\def\begineqn{\begin{eqnarray*}}
\def\endeqn{\end{eqnarray*}}

\def\beginbegin{\begin{equation}}
\def\endbegin{\end{equation}\begin{equation}}
\def\endend{\end{equation}}

\def\beginbib{\begin{thebibliography}{99}}
\def\endbib{\end{thebibliography}}

\def\cl{\centerline}

\date{}

\begin{document}
\begin{center}
{\large{\bf{Effects of Boson-Vacuum Polarization\\
by a Singular Magnetic Vortex}}}\\
{\bf{
Yu. A. Sitenko and A. Yu. Babansky}}\\
Bogolyubov Institute for Theoretical Physics,\\
National Academy of Sciences of Ukraine,\\
Metrologicheskaya ul. l4b, Kiev, 252143 Ukraine 
\vskip 0.7cm

\parbox{11.5cm}{{\bf{Abstract}}-In a space of arbitrary dimensions, 
the effect of an external magnetic field on the vacuum of a quantized
charged scalar field is studied for the field configuration in the 
form of a singular vortex. The zeta-function technique is used to regularize 
ultraviolet divergences. The expression for the effective action is derived.
It is shown that the energy density and current induced in the vacuum decrease 
exponentially at large distances from the vortex. The analytic properties of 
vacuum features as functions of the complex-valued space dimension are 
discussed.}
\end{center}

\vskip 1cm

\section*{\large{1. INTRODUCTION}}

In contrast to what occurs in classical mechanics, 
the motion of charged particles in quantum theory is 
affected by an external electromagnetic field even if the 
region where the field strength is operative is inaccessible
for the particles. For the first time, this was demonstrated for
the example of quantum-mechanical scattering on a flux tube that
is formed by the magnetic lines             +
of force of an external field and which is impenetrable 
for particles undergoing scattering: the scattering cross 
section proved to be a periodic function of the total 
magnetic flux through the tube \cite {AhaR}. This brings about 
the question of whether this configuration, which 
becomes a singular filament (vortex) when the transverse dimensions 
of the flux tube are disregarded, can 
change the properties of the vacuum in a second-quantized theory.

As a matter of fact, this problem has been studied 
for a long time. Although the first steps along these 
lines were made in \cite{Ser}, only in \cite{Sit1} - \cite{Sit4} were reliable 
results obtained for various geometries of space, for 
induced vacuum quantum numbers, and for boundary 
conditions at the locus of a singular magnetic vortex. 
Those studies focused primarily on effects associated 
with the polarization of the vacuum of a quantized 
spinor field. Here, we aim at performing systematic and 
exhaustive inquiries into such effects for a quantized 
scalar field. This case differs from that of a spinor field 
in that some vacuum quantum numbers (charge and 
angular momentum) are not induced in the scalar case;
as a result, the problem reduces to studying the vacuum 
energy and current, as well as quantities that are associated
with them directly (such as. the effective potential generated 
by the vacuum-energy density and the 
strength of the vacuum magnetic field generated by the 
vacuum current) and those that generalize them (zeta 
function). The second distinction between the two 
cases in question is that, in the scalar case, the regularity
 condition can be used for the boundary condition at 
the locus of the vortex. It is the condition that is used 
throughout this study. Finally, the third distinction, 
which follows from the second one, is that the results 
can be immediately generalized to spaces of arbitrary 
dimensions.

\section*{\large{2. QUANTIZATION OF A SCALAR FIELD 
AGAINST THE BACKGROUND OF A SINGULAR 
MAGNETIC VORTEX}}

The operator of a second-quantized complex scalar 
field can be represented in the form
\beginbegin
\Psi(\vx,t)=\Sum{\l}\;\frac{1}{\sqrt{2E_\l}}\left[e^{-iE_\l t}\;\psi_\l (\vx)\,a_\l+
e^{iE_\l t}\;\psi_{-\l}(\vx)\,b_{\l}^{+}\right],
\label{form1}
\endend
where $\l$ is the set of parameters (quantum numbers) 
that specify a state; $E_\l = E_{-\l} >0$ is the energy of a given
state; the symbol $\Sum{\l}$ denotes summation over discrete 
and integration (with a definite measure) over continuous
values of $\l$; $a_\l^{+}$ and $a_\l$ ($b_\l^{+}$ and $b_\l$) are,
respectively. 
the creation and annihilation operators for scalar particles
(antiparticles), the standard commutation relations 
being satisfied for these operators; and the functions 
$\psi_\l ({\bf{x}})$ represent solutions to the time-independent 
Klein-Gordon equation
\beginbegin
\left(-\bfnabla^2 +m^2\right)\psi_\l (\vx)={E_\l}^2 \psi_\l (\vx) ,
\label{form2}
\endend
Here, $\bfnabla$ is the operator of covariant differentiation in a 
static external (background) field. Proceeding in a standard way,
we can show that the vacuum-energy density is given by
\beginbegin
\varepsilon(\vx)=\Sum{\l}\; E_\l \psi_{\l}^{*} (\vx) \psi_\l (\vx)
\label{form3}
\endend
and that the vacuum current has the form
\beginbegin
{\bf{j}}(\vx)=(2i)^{-1} \Sum\l \; {E_\l}^{-1} \Big\{
\psi_{\l}^{*}(\vx) \left[ \bfnabla \psi_{\l}(\vx) \right]
- \left[ \bfnabla \psi_{\l}(\vx) \right]^{*} \psi_{\l}(\vx) \Big\}.
\label{form4}
\endend
If a magnetic field is chosen to represent the background field,
 the covariant derivative takes the form
\beginbegin
\bfnabla={\bfdd}-i{\bf{V}}(\vx),
\label{form5}
\endend
where {\bf{V}}(\vx) is the vector potential of the field in question.
In $d$-dimensional space, the strength of the magnetic field is an
antisymmetric tensor of rank $(d - 2)$
related to the above vector potential by the equation
\beginbegin
B^{\ts{\nu_1 \cdot \cdot \cdot \nu_{d-2}}} (\vx)=
\left[\dd_{\ts{\mu_1}} V_{\ts{\mu_2}}(\vx)\right]
\epsilon^{\ts{\mu_1 \mu_2 \nu_1 \cdot \cdot \cdot \nu_{d-2}}},
\label{form6}
\endend
where $\epsilon^{\ts{\mu_1 \cdot \cdot \cdot \mu_d}}$ is the fully
antisymmetric tensor normalized by the condition
$\epsilon^{1 2 \cdot \cdot \cdot d}=1$.

In this study, we consider the magnetic-field configuration in the form
of a singular vortex represented by a point for $d = 2$, a line for $d = 3$,
and a $(d - 2)$-dimensional hypersurface for $d > 3$; that is, we have
\beginbegin
V_1(\vx)=-\f \frac{x^2}{(x^1)^2+(x^2)^2}, \,\,\,
V_2(\vx)=\f \frac{x^1}{(x^1)^2+(x^2)^2}, \,\,\,
V_\nu(\vx)=0, \,\,\,
\nu=\overline{3,d}\,\, ,
\label{form7}
\endend
\beginbegin
B^{3 \cdot \cdot \cdot d}(\vx)=2\pi\, \f\, \delta(x^1)\, \delta(x^2),
\label{form8}
\endend
where $\f$ is the flux of the vortex in question (in $2\pi$ units).

Let us construct the complete set of solutions to 
equation (\ref{form2}) in the field of the vortex that satisfy the 
regularity condition on the vortex hypersurface. We can 
easily obtain
\beginbegin
\psi_{\ts{kn\vp}}(\vx)=(2\pi)^{\ts{\frac{1-d}{2}}} \Jnf(kr)e^{\ts{in\ugl}}
e^{\ts{i {\bf{px}}_{d-2}}},
\label{form9}
\endend
where
\beginbegin
0<k<\infty, \text n \in \Z, \text -\infty < p_\nu < \infty,\text
\nu=\overline{3,d}\,\,,
\label{form10}
\endend
$J_\omega(u)$ is a Bessel function of order $\omega$,\,\,\,
$r=\sqrt{\left[ (x^1)^2 + (x^2)^2 \right]}$,\,\,
$\ugl=\mbox{arctg}(x^2/x^1)$,\,\,
$\vx_{d-2}=\left(0,0,x^3,...,x^d\right)$ and
$\Z$ is the set of integers. Since the solutions 
given by (9) correspond to a continuous spectrum 
$(E_{kn\vp}=\sqrt{p^2+k^2+m^2}>m)$, they are normalized to a 
delta function:
\beginbegin
\int d^d x \,\,\psi^{*}_{{kn\vp}}(\vx) \psi_{{k'n'{\vp}'}}(\vx) = 
\frac{\delta(k-k')}{k} \delta_{{nn'}}\delta(\vp-{\vp}').
\label{form11}
\endend

Taking all the above into account, we obtain
\beginbegin
\varepsilon(\vx)=(2\pi)^{1-d} \int d^{d-2}p
\integ dk k \left(p^2+k^2+m^2\right)^{{\frac{1}{2}}} \sum_{n \in \Z} \Jnf^2(kr)
\label{form12}
\endend
\begineq
&~&j_{\ugl}(\vx)\equiv r^{-1}\left[ x^1 j_2(\vx) - x^2 j_1(\vx)\right]=
\nonumber\\
&&=(2\pi)^{1-d}\,r^{-1}\int d^{d-2}p
\integ dk k \left(p^2+k^2+m^2\right)^{{-\frac{1}{2}}} \sum_{n \in \Z}(n-\f) \Jnf^2(kr)
\label{form13}\hspace*{1cm}
\endeq
(the remaining components of the vacuum current vanish identically). 
It is not quite correct to define the vacuum-energy density via (\ref{form12})
because the integral involved diverges for $E_{kn\vp}\to\infty$.
All problems associated with regularization of ultraviolet divergences and 
with renormalization can be conveniently solved with 
the aid of a zeta function, which therefore comes to be 
of paramount importance for problems of the type considered here.

\section*{\large{3. ZETA FUNCTION}}

We define the density of the zeta function as [compare with  (\ref{form3})]
\beginbegin
\zeta_{\vx}(s)=\Sum{\l}\; {E_\l}^{-\ts{2s}} \psi_{\l}^{*} (\vx) \psi_\l (\vx).
\label{form14}
\endend
For an external field in the form of a singular magnetic 
vortex, expression (\ref{form14}) takes the form
\beginbegin
\zeta_{\vx}(s)
=(2\pi)^{1-d}\int d^{d-2}p
\integ dk k \left(p^2+k^2+m^2\right)^{-\ts{s}} \sum_{n \in \Z} \Jnf^2(kr).
\label{form15}
\endend
The last expression is well defined only for $\ds{\Re\;s>\frac{d}{2}}$.

Representing the flux of the vortex as the sum of the 
integral ($n'$) and fractional ($F$) parts,
\beginbegin
\f=n'+F, \text n' \in \Z, \text 0 \leq F <1,
\label{form16}
\endend
we find that the zeta-function density (\ref{form15}) depends 
periodically on 0 and that the case of integral flux values ($F=0$)
is equivalent to the flux-free case ($\f=0$). Defining the zeta-function 
density in a free theory (that is, in the absence of vortices) as
\beginbegin
\zeta_{\vx}^{(0)}(s)=\left. \zeta_{\vx}(s) \right|_{F=0},
\label{form17}
\endend
we can easily find that, under the condition $\Re\;s>\frac{d}{2}$
it can be recast into the form
\beginbegin
\zeta_{\vx}^{(0)}(s)
=(2\pi)^{1-d}\int d^{d-2}p
\integ dk k \left(p^2+k^2+m^2\right)^{-\ts{s}}
= \frac{m^{{d-2s}}}{(4\pi)^{\wwd}}\;\frac{\G \left(s-\frac{d}{2}\right)}{\G(s)},
\label{form18}
\endend  
where $\G(z)$ is the Euler gamma function.

For $0<F<1$, the result obtained for the zeta-function density by
performing integration with respect to $\vp$ and summation over $n$ 
in (\ref{form15}) is
\begineq
&&\zeta_{\vx}(s)
= \frac{2}{(4\pi)^{\wwd}}\frac{\G \left(s-\frac{d}{2}+1\right)}{\G(s)}
\integ dk\, k\, \left(k^2+m^2\right)^{{\frac{d}{2}-s-1}}\;\times
\nonumber\\
&&\times\; \int_0^{kr}dy \Big[ J_{F}(y)J_{-1+F}(y)+J_{1-F}(y)J_{-F}(y)\Big],
\label{form19}
\endeq
This expression can be recast into the form
\begineq
&&\zeta_{\vx}(s)
= \frac{r}{(4\pi)^{\wwd}}\frac{\G \left(s-\frac{d}{2}\right)}{\G(s)}\;\times
\nonumber\\
&&\times\;
\integ dk\, \left(k^2+m^2\right)^{{\frac{d}{2}-s}}
\Big[ J_{F}(kr)J_{-1+F}(kr)+J_{1-F}(kr)J_{-F}(kr)\Big].
\label{form20}
\endeq
By using the relation
\beginbegin
\alpha^{-{z}}=\frac{1}{\G(z)}\integ dy\, y^{{z-1}}\, e^{-{\alpha\,y}}, \text \Re\;z>0,
\label{form21}
\endend
we can then reduce expression (\ref{form20}) to the final form
\beginbegin
\zeta_{\vx}(s)=\frac{4\sin(F\pi)}{(4\pi)^{\wwdd}}\,\frac{m^{{d-2s}}}{\G(s)}
\integ du\, e^{-{u}}\, \left[ K_F(u) + K_{1-F}(u) \right]\,
\gamma \left( s-\frac{d}{2},\; \frac{m^2 r^2}{2u} \right),
\label{form22}
\endend
where
\beginbegin
\gamma(z,w)=\int_{{0}}^{{w}} dy\, y^{{z-1}}\, e^{-{y}}
\label{form23}
\endend
is the incomplete gamma function, and $K_\omega(z)$
is the Macdonald function of order $\omega$.

With the aid of the inverse Mellin transformation, 
we find that the kernel of the heat equation is given by

\beginbegin
\langle
\vx \left| \exp\left[ -t\left( -\bfnabla^2+m^2\right)\right] \right| \vx
\rangle
=\frac{1}{2\pi i}\int_{c-i\infty}^{c+i\infty} ds\, t^{-\ts{s}}\, \G(s)\zeta_{\vx}(s),
\text c>\frac{d}{2}.
\label{form24}
\endend
Although expression (\ref{form22}) was obtained under the condition
${\Re\;s>\frac{d}{2}}$,  we can construct an analytic continuation of
this expression to the region ${\Re\;s<\frac{d}{2}}$ by using 
repeatedly the recursion relation
\beginbegin
\gamma(z,w)=
\frac{1}{z}\left[ \gamma(z+1,w)+ w^z e^{-w} \right],
\label{form25}
\endend
As a result, we arrive at the zeta-function density 
expressed in terms of a meromorphic function on the $s$
plane, the products $\G(s)\zeta_{\vx}(s)$ having only simple poles 
on the real axis. The residues at these poles are given by
\beginbegin
\Res_{s=\frac{d}{2}-N+1}\G(s)\zeta_{\vx}(s)=
\frac{1}{(4\pi)^{\wwd}}\frac{(-m^2)^{N-1}}{\G(N)}, \text N=1,2,...\,\, .
\label{form26}
\endend
Shifting the contour of integration in (\ref{form24}) to the left, 
closing it by the corresponding part of a large circle, 
and summing the contributions of the residues, we 
obtain
\beginbegin
\langle
\vx \left| \exp\left[ -t\left( -\bfnabla^2+m^2\right)\right]\right| \vx
\rangle
=\frac{1}{(4\pi t)^{\wwd}}e^{-{m^2 t}}
\Bigg\{ 1+ O \bigg[ \frac{\sqrt{t}}{r}\exp(-\frac{r^2}{t}) \bigg] \Bigg\}.
\label{form27}
\endend
The same expression [$O \bigg[ \frac{\sqrt{t}}{r}\exp(-\frac{r^2}{t}) \bigg]$ 
term apart] is obtained for the kernel of the heat equation in free theory
as well if the function $\zeta^{(0)}_\vx(s)$ (\ref{form18}), which can be 
treated as an analytic continuation to the entire complex 
$s$ plane, is substituted for the function $\zeta_\vx(s)$ (\ref{form22}).
Thus, the functions $\zeta_\vx(s)$ and $\zeta^{(0)}_\vx(s)$ have the same
structure of singularities. Hence, the difference
\beginbegin
\zeta^{ren}_\vx(s)=\zeta_\vx(s)-\zeta^{(0)}_\vx(s)
\label{form28}
\endend
which represents the renormalized zeta-function density, is a holomorphic
function on the complex $s$ plane, and so is the product
$\G(s)\zeta^{ren}_\vx(s)$ . We can easily obtain
\beginbegin
\zeta^{ren}_\vx(s)
=-\frac{4\sin(F\pi)}{(4\pi)^{\wwdd}}\frac{m^{d-2s}}{\G(s)}
\integ du\, e^{-u}\, \left[ K_F(u) + K_{1-F}(u) \right]\,
\G \left( s-\frac{d}{2},\, \frac{m^2 r^2}{2u} \right),
\label{form29}
\endend
where
\beginbegin
\G(z,w)=\G(z)-\gamma(z,w)=\int_{w}^{\infty} dy\, y^{{z-1}}\, e^{-{y}}.
\label{form30}
\endend
Considering that the Macdonald function has the integral representation
\beginbegin
K_\omega(u)=\integ dy\; \ch(\omega y)\; \exp(-u\; \ch y),
\label{form31}
\endend
we eventually obtain
\begineq
\zeta^{ren}_\vx(s)&=&-\frac{16\sin(F\pi)}{(4\pi)^{\wwdd}\G(s)}
\left( \frac{r}{m} \right)^{{s-\frac{d}{2}}}\times
\nonumber\\
&\times& \int_1^\infty \frac{d\ups}{\sqrt{\ups^2-1}}\;\ch \left[ (2F-1)\Arch\ups \right]\;
\ups^{{s-\frac{d}{2}-1}}\;K_{{s-\frac{d}{2}}}(2mr\ups).\;
\label{form32}
\endeq
Thus, the product $\G(s)\zeta^{ren}_\vx(s)$ is a nonpositive function 
of $F$, is symmetric with respect to the substitution $F\to 1-F$, has
a minimum at $F=\frac{1}{2}$, and vanishes at $F=0$. In contrast to the
functions $\zeta^{(0)}_\vx(s)$ (\ref{form18})
and $\zeta_\vx(s)$ (\ref{form22}), the function
$\zeta^{ren}_\vx(s)$ (\ref{form32}) decreases exponentially 
at large distances:
\beginbegin
\zeta^{ren}_\vx(s) =-\frac{\sin(F\pi)}{(4\pi)^{\wwd}\G(s)}e^{-{2mr}}
m^{{\frac{d}{2}-s-1}}r^{{s-\frac{d}{2}-1}}
\left\{
1+ O \left[ (mr)^{-1} \right]
\right\}, \text mr \gg 1. 
\label{form33}
\endend

Instead of (\ref{form21}), we can use the relation
\beginbegin
\alpha^{-{z}}=\frac{2\sin(z\pi)}{\pi}\integ dy\;
\frac{y^{{1-2z}}} {y^2+\alpha},
\text 0<\Re\;z<1,
\label{form34}
\endend
whence it follows that the function $\zeta^{ren}_\vx(s)$ can
alternatively be represented as
\beginbegin
\zeta^{ren}_\vx(s) =-\frac{16\sin(F\pi)}{(4\pi)^{\wwdd}}
\, \frac{r^{{2s-d}}}{\G(s)\G(1-s+\frac{d}{2})}
\int_{mr}^\infty dw\; (w^2-m^2 r^2)^{{\frac{d}{2}-s}}\; K_F(w)K_{1-F}(w).
\label{form35}
\endend
In this representation, the product $\G(s)\zeta^{ren}_\vx(s)$ is a
holomorphic function in the region $\Re\;s<\frac{d}{2}+1$, but it is 
undefined for $\Re\;s>\frac{d}{2}+1$.

In the absence of a vortex, integration of the zeta-function density
with respect to any of the space coordinates obviously leads to a
divergence because $\zeta^{(0)}_\vx(s)$ 
is independent of $\vx$ and because the space being considered is not compact.
In the presence of a vortex, a similar situation arises if we integrate 
the zeta-function density $\zeta_\vx(s)$, which is uniform only in
the direction parallel to the vortex. The situation is different if 
integration is performed with the renormalized zeta-function density
$\zeta^{ren}_\vx(s)$, which decreases exponentially in 
the plane orthogonal to the vortex. It is because of this 
property of the renormalized zeta-function density that 
there exist $s$ values at which the following integral is 
finite:
\beginbegin
\int_{-\infty}^{\infty}dx^1 \int_{-\infty}^{\infty} dx^2 \; \zeta^{ren}_\vx(s)
= - \frac{m^{d-2s-2}}{2(4\pi)^{\frac{d}{2}-1}}
\frac{\G(s-\frac{d}{2}+1)}{\G(s)}\,F(1-F).
\label{form36}
\endend
Here, integration is performed for $\Re\;s>\frac{d}{2}-1$ in the 
case of representation (\ref{form32}) and for  $\frac{d}{2}-1<\Re\;s<\frac{d}{2}+1$
in the case of representation (\ref{form35}). Obviously, the 
result can be continued analytically over the entire 
complex $s$ plane.

\section*{\large{4. EFFECTIVE ACTION 
AND VACUUM ENERGY}}

Going over to the imaginary time $t=-i\tau$, we define 
the effective action in $(d + 1)$-dimensional Euclidean 
spacetime as
\beginbegin
S^{{\rm{eff}}} [{\bf{V}}(\vx)]=\int d\tau \,d^d x\, \Omega(\vx,\tau)=
\ln \det \left( -\dd_\tau^2 - \bfnabla^2+m^2\right)M^{-2},
\label{form37}
\endend
where $M$ is a parameter that has dimensions of mass, 
and $\Omega$ is the effective-action density. In $(d + 1)$-dimensional
space, the effective-action density is expressed in 
terms of the zeta-function density as
\beginbegin
\Omega(\vx,\tau)=-\left. \left[ \frac{d}{ds}\zeta_{\vx,\tau}(s)\right]\right|_{s=0}
- \zeta_{\vx,\tau}(0) \ln M^2.
\label{form38}
\endend
Taking into account (\ref{form18}), (\ref{form22}), (\ref{form28}), and
(\ref{form35}), we find that, in the presence of a singular magnetic vortex, 
expression (\ref{form35}) can be recast into the form
\begineq
&&\Omega(\vx,\tau)=-\frac{4\sin(F\pi)}{(4\pi)^{N+\frac{3}{2}}}m^{2N+1}\times
\nonumber\\
&&\times \integ du\, e^{-u}\left[ K_F(u)+K_{1-F}(u) \right]
\gamma\left(-N-\frac{1}{2},\, \frac{m^2 r^2}{2u}\right),
\,\,d=2N,\,\,\,\,
\label{form39}
\endeq
\begineq
&&\Omega(\vx,\tau)=\frac{m^{2(N+1)}}{(4\pi)^{N+1}} 
\Bigg\{ \frac{(-1)^{N}}{\G(N+2)}
\left[ \ln \frac{M^2}{m^2}+ \psi(N+2)+\gamma \right]
+
\nonumber\\
&&+\frac{\sin(F\pi)}{\pi}
\integ du\, e^{-u}\left[ K_F(u)+K_{1-F}(u) \right]
\G\left(-N-1,\, \frac{m^2 r^2}{2u}\right)\Bigg\} ,
\, d=2N+1,\nonumber\\
\label{form40}
\endeq
In the alternative representation, we have
\begineq
&&\Omega(\vx,\tau) = \frac{1} {(4\pi)^{N+\frac{1}{2}}\G(N+\frac{3}{2})}
\Bigg\{ (-1)^N \pi\, m^{2N+1} +
\nonumber\\
&&+\frac{4\sin(F\pi)}{\pi r^{2N+1}} 
\int_{mr}^\infty dw\, (w^2-m^2 r^2)^{N+\frac{1}{2}}\, K_F(w)K_{1-F}(w)
\Bigg\},\,d=2N ,
\label{form41}
\endeq
\begineq
&&\Omega(\vx,\tau)=\frac{1} {(4\pi)^{N+1}\G(N+2)}
\Bigg\{ (-1)^N \left[\ln \frac{M^2}{m^2}+ \psi(N+2)+\gamma \right]\,m^{2(N+1)}
+
\nonumber\\
&&+\frac{4\sin(F\pi)}{\pi r^{2(N+1)}}
\int_{mr}^\infty dw (w^2-m^2 r^2)^{N+1}\, K_F(w)K_{1-F}(w)
\Bigg\},\,d=2N+1;\,\,
\label{form42}
\endeq
where $\psi(z)=\ds{\frac{d}{dz}\ln\G(z)}$ is the digamma function, and 
$\gamma=-\psi(1)$ is the Euler constant.

The effective potential that is equivalent to the 
renormalized effective-action density is given by
\beginbegin
\Omega^{ren}(\vx,\tau)=
-\left. \left[ \frac{d}{ds}\zeta^{ren}_{\vx,\tau}(s)\right]\right|_{s=0},
\label{form43}
\endend
Equation (\ref{form43}) can be rewritten in an integral form as
\beginbegin
\ln \frac{\det \left( -\dd_\tau^2 - \bfnabla^2+m^2\right)}   
{\det \left( -\dd_\tau^2 - \bfdd^2+m^2\right)}
=\int d\tau \, d^d x\, \Omega^{ren}(\vx,\tau);
\label{form44}
\endend
By considering that the product $\G(s)\zeta^{ren}_{\vx,\tau}(s)$ is
a holomorphic function, it can be shown that the relation
\beginbegin
\zeta^{ren}_{\vx,\tau}(0)=0,
\label{form45}
\endend
must hold. Taking into account the explicit form of the 
renormalized zeta-function density as given by 
(\ref{form32}) or (\ref{form35}), we can easily obtain
\beginbegin
-\left. \left[ \frac{d}{ds}\zeta^{ren}_{\vx,\tau}(s)\right]\right|_{s=0}
= \zeta^{ren}_{\vx}\left(-\frac{1}{2}\right).
\label{form46}
\endend
By comparing equations (\ref{form3}) and (\ref{form14}), we can therefore
naturally define the vacuum-energy density in 
terms of the effective potential as\footnote{Generally, it may turn out that,
at odd values of $d$, the product 
$\G(s)\zeta^{ren}_{\vx,\tau}(s)$ has a simple pole at $s=0$. In this case,
the result corresponding to defining the renormalized vacuum-energy density 
in terms of the effective potential would be different from that for 
the renormalized vacuum-energy density defined in terms of the
function $\zeta^{ren}_{\vx}(-\frac{1}{2})$. It is the situation that arises
when the vacuum energy is induced by an external magnetic field having a 
regular configuration (see \cite{SitR3}).}
\beginbegin
\varepsilon^{ren}(\vx)=\Omega^{ren}(\vx,\tau).
\label{form47}
\endend
As a result, we obtain
\beginbegin
\varepsilon^{ren}(\vx) =\frac{16\sin(F\pi)}{(4\pi)^{\frac{d+3}{2}}}
\left( \frac{m}{r} \right)^{\frac{d+1}{2}}
\int_1^\infty \frac{d\ups}{\sqrt{\ups^2-1}} \ch \left[ (2F-1)\Arch\ups \right]
\ups^{-\frac{d+3}{2}}K_{\frac{d+1}{2}}(2mr\ups),
\label{form48}
\endend
In the alternative representation, we have
\beginbegin
\varepsilon^{ren}(\vx)=\frac{16\sin(F\pi)}{(4\pi)^{\frac{d+3}{2}}}
\frac{r^{-d-1}}{\G\left(\frac{d+3}{2}\right)}
\int_{mr}^\infty dw\, (w^2-m^2 r^2)^{\frac{d+1}{2}} K_F(w)K_{1-F}(w).
\label{form49}
\endend

By using the relations
\beginbegin
\G\left(-N-\frac{1}{2},w\right)=\frac{(-1)^{N}}{\G\left(N+\frac{3}{2}\right)}
\left[
-\pi\, \mbox{erfc}(\sqrt{w})
+e^{-w}\sum_{j=0}^{N}(-1)^{j}\G \left(j+\frac{1}{2}\right) w^{-j-\frac{1}{2}}
\right]
\label{50}
\endend
\beginbegin
\G\left(-N-1,w\right)=\frac{(-1)^{N}}{\G(N+2)}
\left[
-\mbox{E}_{1}(w)+e^{-w}\sum_{j=0}^{N}(-1)^{j}\G(j+1) w^{-j-1} \right],
\label{form51}
\endend
where
\begineqn
\mbox{erfc}(w)=\frac{2}{\sqrt{\pi}}\int_w^{\infty}du\, e^{-u^2}
\endeqn
is the complementary error function, and
\begineqn
\mbox{E}_1(w)=\int_w^{\infty} \frac{du}{u}\, e^{-u}
\endeqn
is the exponential integral (see, for example, \cite{Abra}), we 
find that the maximum value of the renormalized vacuum-energy density
as a function of $F$ is given by
\begineq
&&\left. \varepsilon^{ren}(\vx) \right|_{F=\frac{1}{2}}=
\frac{2 m^{2N+1}} {(4\pi)^{N+\frac{1}{2}}}
\frac{(-1)^{N}}{\G\left(N+\frac{3}{2}\right)}
\Bigg\{
-\frac{\pi}{2}+ \pi mr \Big[ K_0(2mr)L_{-1}(2mr)+
\nonumber\\
&&+K_1(2mr)L_0(2mr) \Big]+\frac{1}{\sqrt{\pi}}
\sum_{l=0}^{N}(-1)^{l} \G\left(l+\frac{1}{2}\right) (mr)^{-l} K_{l+1}(2mr)
\Bigg\},\;d=2N,
\nonumber\\ \label{form52}
\endeq
\begineq
&&\left. \varepsilon^{ren}(\vx) \right|_{F=\frac{1}{2}}=
\frac{2 m^{2(N+1)}} {(4\pi)^{N+1}} 
\frac{(-1)^{N}}{\G\left(N+2\right)}
\Bigg\{
-\mbox{E}_1(2mr)+
\nonumber\\
&&+e^{-2mr}
\sum_{l=0}^{N}(-1)^{l} \G\left(l+1\right)
\sum_{n=0}^{l+1}
\frac{\G(l+n+2) (mr)^{-l-n-1}} {2^{2n+1}\G(n+1)\G(l-n+2)} 
\Bigg\},\, d=2N+1,\hspace{1cm}
\label{form53}
\endeq
where $L_{\omega}(u)$ is the modified Struve function of order
$\omega$ \cite{Abra}. In particular, we have
\begineq
\left. \varepsilon^{ren}(\vx) \right|_{F=\frac{1}{2}}&=&
\frac{m^3} {3\pi^2}
\Bigg\{
\frac{\pi}{2}- \pi mr \left[ K_0(2mr)L_{-1}(2mr)+K_1(2mr)L_0(2mr) \right]-
\nonumber\\
&-& K_1(2mr)+(2mr)^{-1}K_2(2mr)
\Bigg\}, \, d=2\, ,
\label{form54}
\endeq
\beginbegin
\left. \varepsilon^{ren}(\vx) \right|_{F=\frac{1}{2}}=
\frac{m^{4}} {16\pi^2}
\left\{
\mbox{E}_1(2mr)+\frac{e^{-2mr}}{2mr}
\left[-1+\frac{1}{2mr}+\frac{3}{2(mr)^2}+\frac{3}{4(mr)^3}\right]
\right\},\, d=3.
\label{form55}
\endend

To conclude this section, we present the asymptotic 
expressions for the renormalized vacuum-energy density at small and
large distances from the vortex. We have
\beginbegin
\varepsilon^{ren}(\vx)=\frac{4\sin(F\pi)}{(4\pi)^{\frac{d}{2}+1}}
\frac{\G\left(\frac{d+1}{2}+F\right)\G\left(\frac{d+1}{2}+1-F\right)}
{(d+1)\G\left(\frac{d}{2}+1\right)}\, r^{-d-1} 
\left\{ 1+O\left[(mr)^2\right]\right\}, \, mr\ll 1,
\label{form56}
\endbegin
\varepsilon^{ren}(\vx)=\frac{\sin(F\pi)}{(4\pi)^{\frac{d+1}{2}}}\,
e^{-2mr}\, m^{\frac{d-1}{2}}\, r^{-\frac{d+3}{2}} 
\left\{ 1+O\left[(mr)^{-1}\right]\right\}, \, mr\gg 1.
\label{form57}
\endend

\section*{\large{5. VACUUM CURRENT AND VACUUM 
MAGNETIC FIELD}}

Performing integration with respect to $p$ in (\ref{form13}) and 
using relation (\ref{form21}), we find that the vacuum current is 
given by\footnote{As a matter of fact, an analytic continuation in 
the complex-valued variable $d$ is performed here from the region
$\Re\,d<3$ to the region $\Re\,d>3$ (compare with the presentation in
Section 3). We then arrive at the quantity expressed in (\ref{form58})
in terms of a function that is holomorphic over the entire complex $d$ plane.}
\beginbegin
j_{\ugl}(\vx)=\frac{4}{r\,(4\pi)^{\frac{d+1}{2}}}
\integ dy\, y^{\frac{1-d}{2}}\, e^{-m^2 y}
\integ dk\, k\, e^{-k^2 y} \sum_{n \in \Z}(n-\f)\Jnf^2(kr).
\label{form58}
\endend
Performing summation over $n$ in the expression on the 
right-hand side of (\ref{form58}), we obtain
\begineq
&~&j_{\ugl}(\vx)=\frac{2}{(4\pi)^{\frac{d+1}{2}}}
\integ dy\, y^{\frac{1-d}{2}}\, e^{-m^2 y}
\integ dk \,k^2 e^{-k^2 y} \Bigg\{kr\Big[ J_{1-F}^2(kr)+ J_{-F}^2(kr)-\vspace{2cm}
\nonumber\\
&~& - J_{F}^2(kr)- J_{-1+F}^2(kr) \Big] 
+(2F-1)
 \Big[ J_{1-F}(kr)J_{-F}(kr)+J_{F}(kr)J_{-1+F}(kr)\Big] \Bigg\}.
\nonumber\\
\label{form59}
\endeq
Following integration with respect to $k$ and the substitution
$\ds{\frac{r^2}{2y}=u}$, we reduce the expression for $j_{\ugl}(\vx)$ to the
form
\begineq
j_{\ugl}(\vx)=\frac{2\sin(F\pi)}{(2\pi)^{\frac{d+3}{2}}}\,r^{-d}
\integ du\, u^{\frac{d-1}{2}}\exp\left(-u-\frac{m^2 r^2}{2u}\right)
\Big[K_F(u)-K_{1-F}(u) \Big],
\label{form60}
\endeq
With the aid of the integral representation (\ref{form31}) for the 
Macdonald function, we then obtain
\beginbegin
j_{\ugl}(\vx)=\frac{32\sin(F\pi)}{(4\pi)^{\frac{d+3}{2}}}\,
m^{\frac{d+1}{2}}\,r^{-\frac{d-1}{2}}
\int_1^\infty d\ups \; \sh \left[ (2F-1)\Arch\ups \right]\;
\ups^{-\frac{d+1}{2}}\;K_{\frac{d+1}{2}}(2mr\ups).\;
\label{form61}
\endend

If relation (\ref{form34}) is used instead of (\ref{form21}), we arrive at
\begineq
&&j_{\ugl}(\vx)=
\frac{32\sin(F \pi)}{(4\pi)^{\frac{d+3}{2}}}\frac{r^{-d}}{\G(\frac{d-1}{2})}
\int_{mr}^\infty dw\, w^2\,
(w^2-m^2 r^2)^{\frac{d-3}{2}}\,\times
\nonumber\\
&&\times\,
\Bigg\{
w\Big[K_{1-F}^2(w)-K_{F}^2(w)\Big]
+\Big(2F-1\Big) K_F(w)K_{1-F}(w)
\Bigg\};
\label{form62}
\endeq
In this representation, the vacuum current is a holomorphic function of
$d$ in the region $\Re\, d>1$, but it is not defined for
$\Re\, d<1$.

In the case of spaces of lower dimensions ($d$ = 2, 3, 
4, and 5), it can be shown by transforming the integral 
in equation (\ref{form60}) that the corresponding expressions for 
the vacuum current are
\begineq
&&j_{\ugl}(\vx)=\frac{\sin(F\pi)}{4\pi^2 r^2} \Big(F-\frac{1}{2}\Big)
\Bigg\{
-4\Big[ (F-\frac{1}{2})^2+m^2 r^2\Big]
\int_{2mr}^{\infty} \frac{du}{u} K_{2F-1}(u)+
\nonumber\\
&&+mr\Big[K_{2F}(2mr)+K_{2(1-F)}(2mr)\Big] \Bigg\}, \;\;\; d=2,
\label{form63}
\endeq
\begineq
&&j_{\ugl}(\vx)=\frac{\sin(F\pi)}{6 \pi^3}\frac{m}{r^2}
\Bigg\{
\Big[ F(F-\frac{1}{2})+ m^2 r^2\Big]\,mr\,K_{F}^2(mr)-
\nonumber\\
&&
-\Big[ (1-F)(\frac{1}{2}-F)+m^2 r^2\Big]\,mr\,K_{1-F}^2(mr)+
\nonumber\\
&&
+2\Big[{F(1-F)}-m^2 r^2\Big]
(F-\frac{1}{2})K_{F}(mr)K_{1-F}(mr)
\Bigg\},\;\; d=3,
\label{form64}
\endeq
\begineq
&&j_{\ugl}(\vx)=\frac{\sin(F\pi)}{32\pi^3 r^4} \Big(F-\frac{1}{2}\Big)
\Bigg(
4\bigg\{(F-\frac{1}{2})^2 \Big[ (F-\frac{1}{2})^2
-1+2m^2 r^2 \Big] + m^4 r^4 \bigg\}\times
\nonumber\\
&&\times
\int_{2mr}^{\infty} \frac{du}{u} K_{2F-1}(u)+2m^2 r^2 K_{2F-1}(2mr)-
\nonumber\\
&&-\Big[(F-\frac{1}{2})^2-1+m^2 r^2\Big]\,mr\,\Big[K_{2F}(2mr)+K_{2(1-F)}(2mr)\Big]
\Bigg), \;\; d=4,
\label{form65}
\endeq
\begineq
&&j_{\ugl}(\vx)=\frac{\sin(F\pi)}{60 \pi^4}\frac{m}{r^4}
\Bigg(
\bigg\{ (F-\frac{1}{2})\Big[ (1+F)F(2-F)-(2F-\frac{1}{2})m^2 r^2\Big]-
\nonumber\\
&&- m^4 r^4 \bigg\}\,mr\,K_{F}^2(mr)
-\bigg\{ (\frac{1}{2}-F)\Big[ (1+F)(1-F)(2-F)-(\frac{3}{2}-2F)m^2 r^2\Big]-
\nonumber\\
&&- m^4 r^4 \bigg\}\,mr\,K_{1-F}^2(mr)
+2\bigg\{ (1+F)F(1-F)(2-F)+
\nonumber\\
&&+\Big[1-2F(1-F)\Big]m^2 r^2
+ m^4 r^4\bigg\}(F-\frac{1}{2})K_{F}(mr)K_{1-F}(mr) \Bigg),\;\; d=5,
\label{form66}
\endeq
The expression on the right-hand side of (\ref{form64}) was first 
obtained in \cite{Ser}.\footnote{
The results presented in \cite{Ser} for the vacuum energy density at 
$d = 3$ are self-contradictory and erroneous.}
The integral representation (\ref{form60}) for 
the case of $d=3$ was also presented in \cite{Jor}. 

The asymptotic expressions for the vacuum current at 
small and large distances from the vortex are given by
\begineq
j_{\ugl}(\vx)=\frac{4\sin(F\pi)}{(4\pi)^{\frac{d}{2}+1}}
\left(F-\frac{1}{2} \right)
\frac{\G\left(\frac{d-1}{2}+F\right)\G\left(\frac{d-1}{2}+1-F\right)}
{\G\left(\frac{d}{2}+1\right)}\, r^{-d} 
\left\{ 1+O\left[(mr)^2\right]\right\},
\nonumber\\
mr\ll 1,\;\;
\label{form67}
\endeq
\beginbegin
j_{\ugl}(\vx)=\frac{2\sin(F\pi)}{(4\pi)^{\frac{d+1}{2}}}\left(F-\frac{1}{2} \right)
e^{-2mr} m^{\frac{d-3}{2}} r^{-\frac{d+3}{2}} 
\left\{ 1+O\left[(mr)^{-1}\right]\right\}, \, mr\gg 1.
\label{form68}
\endend
To conclude this section, we note that, according to 
the Maxwell equation
\beginbegin
\dd_r B^{3...d}_{(I)}(\vx)=-e^2 j_{\ugl}(\vx),
\label{form69}
\endend
where $e$ is the coupling constant having dimensions of 
$m^{{\frac{3-d}{2}}}$ in $(d+1)$-dimensional spacetime, the magnetic 
field of strength
\beginbegin
B^{3...d}_{(I)}(\vx)=e^2 \int_r^\infty dr\, j_{\ugl}(\vx).
\label{form70}
\endend
is induced in the vacuum. Using relation (\ref{form61}) or (\ref{form62})
for $j_{\ugl}(\vx)$, we obtain
\beginbegin
B^{3...d}_{(I)}(\vx)=
\frac{16\,e^2\,\sin(F\pi)}{(4\pi)^{\frac{d+3}{2}}}
\left( \frac{m}{r}\right)^{\frac{d-1}{2}}
\int_1^\infty d\ups \; \sh \left[ (2F-1)\Arch\ups \right]\;
\ups^{-\frac{d+3}{2}}\;K_{\frac{d-1}{2}}(2mr\ups),\;
\label{form71}
\endend
In the alternative representation, we have
\begineq
&&B^{3...d}_{(I)}(\vx)=
\frac{16\,e^2\,\sin(F\pi)}{(4\pi)^{\frac{d+3}{2}}}\frac{r^{1-d}}{\G(\frac{d+1}{2})}
\int_{mr}^\infty dw\, 
(w^2-m^2 r^2)^{\frac{d-1}{2}}\,\times
\nonumber\\
&&
\times\,
\bigg\{
w\Big[K_{1-F}^2(w)-K_{F}^2(w)\Big]+
\Big(2F-1\Big) K_F(w)K_{1-F}(w)
\bigg\}.
\label{form72}
\endeq
The total flux of the vacuum magnetic field (in $2\pi$
units),
\beginbegin
\Phi^{(I)}=\integ dr\, r B^{3...d}_{(I)}(\vx)
\label{form73}
\endend
diverges for $d \ge 3$ and is finite in all other cases. Specifically,
we have
\begineq
\Phi^{(I)}=\frac{2\,e^2\, m^{d-3}}{3(4\pi)^{\frac{d+1}{2}}}
\G\left(\frac{3-d}{2}\right) F\left(1-F\right)\left(F-\frac{1}{2}\right).
\label{form74}
\endeq
It should be emphasized that the last expression can be 
continued analytically from the holomorphicity region 
$\Re\, d<3$ to the entire complex $d$ plane. The result proves 
to be finite at even real values of $d$.

\section*{\cl{6. CONCLUSION}}

   Effects associated with boson-vacuum polarization 
by an external field in the form of a singular magnetic 
vortex have been studied comprehensively. Integral 
representations have been obtained for the zeta-function
density [equation (\ref{form32}) or (\ref{form35})] for the effective
action density [equations (\ref{form39}) and (\ref{form40}) or (\ref{form41})
and (\ref{form42})], for the vacuum-energy density [equation (\ref{form48}) or 
(\ref{form49})], for the vacuum current [equation (\ref{form61}) or (\ref{form62})], 
and for the strength of the vacuum magnetic field
[equation (\ref{form71}) or (\ref{form72})]. Remarkably, our results concerning
the zeta function, energy, current, and magnetic-field strength admit
analytic continuation in the space dimension $d$,
the representations given by (\ref{form32}), 
(\ref{form48}), (\ref{form61}), and (\ref{form71}) being implemented in terms
of functions that are holomorphic in the entire complex 
$d$ planer The global features of the vacuum have also 
been determined [see equations (\ref{form36}) and (\ref{form74})]; of these, 
only the total flux of the vacuum magnetic field is finite 
at the single value of $d= 2$, although expression (\ref{form36}) at 
$s=-\frac{1}{2}$ and expression (\ref{form74}) treated in the sense of an 
analytic continuation are finite at all even values of $d$.

Our results depend periodically, with a period equal 
to unity, on the flux $\f$ of the vortex. At nonintegral values of $\f$,
the vacuum-energy density is a positive function
(convex for $\Re\, d>-1$) of the fractional part $F$ of the 
flux of the vortex, is symmetric with respect to the substitution
$F\to 1-F$, and has a maximum at $F=\frac{1}{2}$. At 
the same time, the vacuum current vanishes at $F=\frac{1}{2}$, 
is a negative function (concave for $\Re\, d>1$) in the interval
$0<F<\frac{1}{2}$, and is a positive function (convex for 
$\Re\, d>1$) in the interval $\frac{1}{2}<F<1$, the minimum and 
the maximum being located symmetrically with respect 
to the point $F=\frac{1}{2}$. Accordingly, the vacuum-energy 
density is even under charge conjugation, while the 
vacuum current is odd. It has been shown that the local 
features of the vacuum decrease exponentially at large 
distances from the vortex'[see equations 
(\ref{form33}), (\ref{form57}) and (\ref{form68}).

To conclude our brief summary of the results, we 
note that the formulas for the vacuum-energy density 
become especially simple in the case of a massless scalar field
[see equations (\ref{form56}) and (\ref{form67})].

\section*{\cl{ACKNOWLEDGEMENTS}}

This work was supported by the State Foundation 
for Basic Research of Ukraine (project no. 2.4/320) and 
by the Swiss National Science Foundation (grant 
no. CEEC/NIS/96-98/7 IP 051219).

\newpage
\beginbib
\cl{{\large\bf{REFERENCES}}}
\bibitem{AhaR}  Aharonov, Y. and Bohm, D., Phys. Rev., 1959. vol. 115, p. 485.
\bibitem{Ser} Serebryanyi, E.M., Teor. Mat. Fiz., 1985, vol. 64. p. 299.
\bibitem{Sit1} Sitenko,Yu.A., Yad. Fit., 1988, vol. 47, p. 292.
\bibitem{Sit2} 4. Sitenko, Yu.A., Nucl. Phys. B. 1990, vol. 342, p. 655;\\
Phys. Lett. B, 1991. vol. 253, p. 138.
\bibitem{Jor} Gornicki, P., Ann. Phys. (N.Y.), 1990, vol. 202, p. 271.
\bibitem{Sit3}  Sitenko, Yu.A., Phys. Lett. B, 1996, vol. 387, p. 334.
\bibitem{SitR1} Sitenko, Yu.A. and Rakityansky. D.G., Yad. Fiz; 1997,\\ 
vol. 60, pp. 308, 320 [Phys. At. Nucl. (Engl. Transl.), vol. 60. pp. 247, 258].
\bibitem{SitR2} Sitenko, Yu.A. and Rakityansky. D.G., Yad. Fit., 1997, 
vol. 60, p. 1643\\
{[Phys. At. Nucl. (Engl. Transl.), vol. 60, p. 1497]. }
\bibitem{Sit4} Sitenko, Yu.A., Yad. Fit., 1997, vol. 60, p. 2285\\
{[Phys. At. Nucl. (Engl. Transl.), vol. 60, p. 2102].}
\bibitem{SitR3} Sitenko, Yu.A. and Rakityansky, D.G., Yad. Fiz., 1998, 
vol. 61, p. 876\\
{[Phys. At. Nucl. (Engl. Transl.), vol. 61, p. 790].}
\bibitem{Abra} Handbook of Mathematical Functions, Abramowitz, M. 
and Stegun, LA., Eds., New York: Dover, 1965.
\endbib

\end{document}